\documentclass[epj,draft]{svjour}
\newcommand{\p}[1]{(\ref{#1})}
\newcommand{\lab}{\label}
\usepackage{amssymb,latexsym}
\usepackage{amsmath}
\begin{document}

\title{Application of Bogolyubov's  approach to the derivation of  kinetic
equations for dissipative systems}
\author{
I. Goldhirsch\inst{1}\,,
A.~S. Peletminskii\inst{2}
S.~V. Peletminskii\inst{2},
A.~I. Sokolovsky\inst{3}
}                     
%
%
\institute{ Department of Fluid Mechanics and Heat Transfer,
Faculty of Engineering, Tel-Aviv University, Ramat-Aviv, Tel-Aviv
69978, Israel \and Akhiezer Institute for Theoretical Physics,
National Science Centre "Kharkov Institute of Physics and
Technology", Kharkov 61108, Ukraine  \and Department of
Theoretical Physics, Dnepropetrovsk National University,
Dnepropetrovsk 49010, Ukraine }
\date{Received: date / Revised version: date}
%
\abstract {The main goal of the present article is to extend  the Bogolyubov method for deriving
kinetic equations  to dissipative many-body systems. The basic conjecture underlying the Bogolyubov
approach is the functional hypothesis, according to which, the many-particle distribution functions
are assumed to be functionals of the one-particle distribution function on kinetic time scales.
Another ingredient in the Bogolyubov approach is the principle of the spatial weakening of
correlations, which reflects statistical independence of physical values at distant spatial points.
One can consider it as a reasonable mixing property of many-particle distribution functions. The
motivation behind the generalization of Bogolyubov's approach to (classical) many-body dissipative
systems is the wish to describe the dynamics of granular systems, in particular granular fluids. To
this end we first define a general dissipative fluid through a dissipation function, thereby
generalizing the commonly employed models for granular fluids. Using the Bogolyubov functional
hypothesis we show how a reduction of the pertinent BBGKY hierarchy can be achieved. The method is
then employed to cases which can be treated perturbatively, such as those in which the interactions
are weak or the dissipation is small or the particle density is small. Kinetic descriptions are
obtained in all of these limiting cases. As a test case, we show that the Bogolyubov method begets
the now standard inelastic Boltzmann equation for dilute monodisperse collections  of spheres whose
collisions are characterized by a fixed coefficient of normal restitution. Possible further
applications and implications are discussed.
} 

\maketitle

\section{Introduction}\label{Introduction}
The dynamics of granular systems in general and granular gases in particular is of much current
interest \cite{IGreview}. Except for the physical  dimensions of typical macroscopic grains, the
main difference between granular and molecular many-body systems is the dissipative nature of the
interactions in the former. This property has far reaching consequences, many of which are a-priori
counterintuitive.

While much of the theory of granular solids and quasi-static granular flows is of phenomenological
nature, granular gases seem to be well described by kinetic theory  \cite{IGreview},  i.e. the
Boltzmann \cite{Sela_Gold,Dufty}, or Enskog-Boltzmann equations \cite{Gol'd_Sok,densegrangases},
with the possible exception of strongly inelastic systems. Attempts to go beyond the Boltzmann
level of description  are noted \cite{GKIG,GK2,vannoije}, but so far these directions have not been
fully exploited.

The Bogolyubov method for deriving kinetic equations for many-body systems \cite{Bog,Bogol} is
based on an assumption known as the functional hypothesis. According to it, for long times
$t\gg\tau_0$, many-particle distribution functions become functionals of the corresponding
one-particle distribution function. Characteristic time $\tau_{0}$ is of the order of the typical
duration of a collision (for hard sphere collisions, $\tau_{0}$ is the time in which a particle
traverses a distance that equals its diameter). The Bogolyubov functional hypothesis can be
considered as a generalization of the Chapman-Enskog method for deriving hydrodynamic equations on
the basis of kinetic equation. It should also be emphasized that the term ``functional hypothesis"
is not, in some measure, adequate because this statement was proved for some important cases (see,
e.g., \cite{Akh_Pel}). In fact, the first proof of the functional hypothesis was given by Gilbert
in discussion of solutions of the Boltzmann equation.

An important component of the Bogolyubov method is the principle of spatial correlations weakening,
which reflects statistical independence of physical values at distant spatial points. One can
consider it as a reasonable mixing property of many-particle distribution functions. The impact of
the Bogolyubov approach on kinetic theory is described in \cite{Cohen}, and a detailed exposition
of his ideas in this field and some of their applications can be found in the monograph
\cite{Akh_Pel}, whose techniques we generalize here to render them applicable to dissipative
systems.

The present article has two main goals. The first is to present a (rather straightforward)
generalization of the standard models of granular gas collisions by introducing a dissipation
function in conjunction with a Hamiltonian formulation of classical mechanics for dissipative
systems. The use of dissipation functions is of course not new, but their application to granular
systems seems to be novel. On the basis of this formulation we develop a BBGKY hierarchy for
dissipative systems. A BBGKY hierarchy for a system of hard inelastically colliding spheres, which
is based on a pseudo-Liouville equation, is presented e.g., in \cite{vannoije,Dufty-1}. However, in
this approach there is a problem of adequacy of description of many-body dynamics by using a
pseudo-Liouville equation.

The second goal of this article is to implement the Bogolyubov method to the derivation of kinetic
equations to the case of dissipative many-body classical systems. As mentioned, this approach is
based on the functional hypothesis. This conjecture seems to be borne out by all studied
nonequilibrium systems we are aware of. In some non-trivial cases, such as the properties of
non-equilibrium steady states, it was shown to successfully reproduce results obtained by other
methods \cite{SokOpp}. Moreover, the Bogolyubov method enables to study, for example, the problem
of convergence and non-analyticity arising in a perturbation theory \cite{Sok}. Whether in the
realm of granular systems it will yield novel results, which the commonly used methods are
incapable of producing, is at present unclear. However, given the difficulties one faces when
studying granular systems we believe it is important to explore the possibilities afforded by an
alternate formulation. Therefore, this article is devoted to the exposition of the Bogolyubov
formulation of the kinetics of dissipative gases. As a simple test, it is shown that in the limit
of a dilute collections of monodisperse spheres interacting by collisions characterized by a fixed
coefficient of normal restitution,  the present formulation reproduces the corresponding
(inelastic) Boltzmann equation. Other cases, which can be treated perturbatively are presented
below.

The structure of this paper is as follows. In section \ref{formulation} we formulate a dissipative
dynamics on the basis of Hamilton equations of motion and dissipation function. Then we derive the
corresponding Liouville equation and BBGKY hierarchy and formulate the Bogolyubov principle of
spatial correlation weakening for many-particle distribution functions. Section \ref{weak}
introduces the Bogolyubov functional hypothesis and boundary condition as necessary concept for
solving BBGKY hierarchy. It also formulates the basic equations for deriving kinetic descriptions
in different limiting cases. In particular, it is shown how a kinetic equation in the limit of weak
interactions can be derived. Section \ref{dilute} deals with kinetic theory in the low density
limit. Here a generalized Boltzmann equation for gases with dissipative interactions is obtained.
Section \ref{weakdisskin} is devoted to a derivation of a generalized Boltzmann equation for the
case of weak dissipation. At the end of this section a sketch of the theory of homogeneous cooling
states is presented. Section \ref{spheres} discusses the connection of the proposed kinetic theory
of gases in the presence of dissipative interaction with the Bolztmann equation for inelastic rigid
spheres. Finally, Section \ref{conclusion} comprises a brief summary and outlook.

\section{Formulation, the  Liouville equation and the BBGKY
hierarchy}\label{formulation}

Consider a system composed of $N$ identical classical particles of
mass $m$ each.  Their reversible interactions are assumed to be
derivable from a Hamiltonian, $H$, and their dissipative
interactions are assumed to be determined by a dissipation
function, $R$.  Both $H$ and $R$ are assumed to  depend on the
spatial  coordinates of the centers of mass of the particles, $\{
{\bf x}_{i};\  1\leqslant i \leqslant N \}$ and the respective
momenta  $\{ \vec p_i;\ 1\leqslant i \leqslant N \} $.   The
generalized Hamilton equations are given by:
\begin{equation} \lab{eq:2.1}
\dot p_{in}=-{\partial H\over\partial
x_{in}}-{\partial R\over\partial p_{in}},\qquad \dot
x_{in}={\partial H\over\partial p_{in}},
\end{equation}
where we assume for simplicity that the particles experience only
binary interactions:
\begin{equation}\lab{eq:2.2}
H=H_{0}+V=\sum_{1\leqslant i\leqslant N}{{p}_{i}^{2}\over
2m}+\sum_{1\leqslant i<j\leqslant N} V_{ij}, \quad V_{ij}\equiv
V(\vec x_{ij}),
\end{equation}
and where $H_0$ denotes the kinetic part of the Hamiltonian. The
dissipation function $R$ is taken to depend on the coordinate
and momentum differences (to preserve Galilean invariance). For
sake of simplicity it assumed to equal a sum of two-particle
interactions:
\begin{equation}\lab{eq:2.3}
R=\sum_{1\leqslant i<j\leqslant N}R_{ij}, \qquad R_{ij}\equiv
R(\vec x_{ij},\vec p_{ij}),
\end{equation}
where ${\bf x}_{ij}\equiv {\bf x}_{i}-{\bf x}_{j}$ and ${\bf p}_{ij} \equiv {\bf p}_{i}-{\bf
p}_{j}$. As the entity  $R_{ij}$ is a scalar, it depends on  $\vec p^2_{ij}$, $\vec x^2_{ij}$ and
${\bf p}_{ij}{\bf x}_{ij}$. The resulting equations of motion are invariant under Galilean
transformations, as they should, and conserve momentum but not energy. The force exerted by
particle $j$ on particle $i$ is defined as
\begin{equation}\lab{eq:2.4}
{F}_{ij,n}\equiv{F_n}({\vec x}_{ij},{\vec p}_{ij})=-{\partial
V_{ij}\over\partial{x}_{in}}-{\partial
R_{ij}\over\partial{p}_{in}}
\end{equation}
The time derivative of total energy of
the system is given by
\begin{equation}\lab{eq:2.4'}
{dH\over dt}=-\sum_{1\leqslant i\leqslant N}{\partial
R\over\partial  p_{in}}{\partial H\over\partial p_{in}}=
-\sum_{1\leqslant i\leqslant N}{p_{in}\over m}
{\partial R\over\partial p_{in}}.
\end{equation}

We shall assume that the nature of dissipation is associated with friction of macroscopic
particles. Therefore, as a model dissipation function, we take $R_{ij}$ as follows \cite{Landau}:
\begin{equation}\lab{eq:2.4''}
R_{ij}={1\over 2}\gamma({\bf x}_{ij}){\bf p}_{ij}^2,
\end{equation}
moreover, $\gamma(\vec x_{ij})$=0 when $|\vec x_{ij}|
\gtrsim r_0$, where $r_{0}$ is the radius of dissipative
interaction, i.e. the dissipative force acts at the moment
of contact of particles. According to (\ref{eq:2.4'}),
(\ref{eq:2.4''}), one finds
$$
{dH\over dt}=-{2\over m}\sum_{1\leqslant i<j\leqslant N}R_{ij}<0.
$$
Since $\gamma({\bf x}_{ij})>0$ (see \cite{Landau}), the energy
dissipation occurs.

As a precursor to the derivation of the corresponding Liouville
equation for the system described by Eqs. \p{eq:2.1}, we first
present some rather well known results concerning general systems
of ordinary differential equations of the form:
\begin{equation} \lab{eq:2.5}
\dot{x}_{i}(t)=h_{i}(x_{1}(t),...,x_{N}(t)), \qquad 1\leqslant i\leqslant N.
\end{equation}
In particular, if $h_{i}$ is a random field, then Eqs. (\ref{eq:2.5}) can be used to derive the kinetic description of stochastic systems \cite{LaskPelPr}.
Denote by $X_i(t,x)$ the solution of the Cauchy problem of this equation with initial condition
$x\equiv (x_1,x_2,...,x_N)$ ($X_i(x,0)\equiv x_i$). It is well known that Eqs.~\p{eq:2.5} admit the
following formal solution:
\begin{equation}\lab{eq:2.6}
X_{i}(t,x)=e^{t \Lambda(x)}x_{i},
\end{equation}
where
\begin{equation}\lab{eq:2.6'}
\Lambda(x)=\sum_{1\leqslant i\leqslant
N}h_{i}(x){\partial\over\partial x_{i}}.
\end{equation}
When the evolution operator $e^{t\Lambda(x)}$ acts on an arbitrary function $\varphi(x)$ the result
is as follows:
\begin{equation}\lab{eq:2.7}
e^{\tau{\Lambda}(x)}\varphi(x)= \varphi(e^{\tau{\Lambda}(x)}x).
\end{equation}
Notice that since Eqs. (\ref{eq:2.5}) are autonomous, the solution (\ref{eq:2.6}) can be inverted:
\begin{equation}\lab{eq:2.8}
x'_i\equiv X_i(t,x) \Rightarrow  x=X(-t,x').
\end{equation}

Define ${\cal D}(x,0)$ to denote the probability distribution of the initial conditions $x$ (see
Eqs. (\ref{eq:2.5})). Normalization requires that
\begin{equation}\label{8.5}
\int dx {\cal D}(x,0)=1\quad (dx\equiv dx_1...dx_N).
\end{equation}
The distribution function at time $t$ is, therefore, given by
\begin{equation}\lab{eq:2.9}
{\cal D}(x,t)=\int dx'{\cal D}(x',0)\prod_{1\leqslant i\leqslant
N}\delta(x_{i}-X_{i}(t,x')).
\end{equation}
Changing the integration variables from $x'$ to $y=X(t,x')$ and using the relation \p{eq:2.8}, one
obtains
\begin{equation}\lab{eq:2.10}
{\cal D}(x,t)=I(x,t){\cal D}(X(-t,x),0),
\end{equation}
where
\begin{equation} \lab{eq:2.11}
I(x,t)=\bigg\vert{\partial X(-t,x)\over\partial x}\bigg\vert
\end{equation}
is the Jacobian of the transformation $x\to X(-t,x)$.

Following \p{eq:2.6}-\p{eq:2.7} and expression \p{eq:2.10}, it can be shown that the distribution
function ${\cal D}(x,t)$ satisfies the equation:
\begin{equation}\lab{eq:2.13}
{\partial{\cal D}\over\partial t}=\left({\partial I\over\partial t}-I\sum_{1\leqslant i\leqslant
N}h_{i}{\partial\over\partial x_{i}}\right)I^{-1}{\cal D}.
\end{equation}
The equation of motion for the Jacobian $I(x,t)$ has the form
\begin{equation}\lab{eq:2.14}
{\partial I(x,t)\over\partial t}+\tilde \Lambda (x)I(x,t)=0,
\qquad I(x,0)=1,
\end{equation}
with operator $\tilde \Lambda (x)$ defined by
\begin{equation}\lab{eq:2.16'} \tilde \Lambda
(x)\varphi (x)\equiv \sum_{1\leqslant i\leqslant N}{\partial\over\partial
x_{i}}\left(h_{i}(x)\varphi(x)\right).
\end{equation}
In deriving \p{eq:2.14} we have employed the fact that $\int dx{\partial{\cal D}/\partial t}=0$
(see \p{eq:2.9}) and that this relation should hold for any allowed initial distribution function.
Upon elimination of $\partial I/\partial t$ on the right-hand side of \p{eq:2.13}, one obtains
\begin{equation}\lab{eq:2.15}
{\partial{\cal D}(x,t)\over\partial t}+\tilde\Lambda(x) {\cal
D}(x,t)=0.
\end{equation}
A comparison of \p{eq:2.10} with the solution of Eq. \p{eq:2.15} gives the following operator
relation:
\begin{equation}\lab{eq:2.17}
e^{-t\tilde\Lambda(x)}=I(x,t)e^{-t\Lambda(x)}.
\end{equation}


At this stage we return to the dynamical model \p{eq:2.1}--\p{eq:2.3}. In this case, the
corresponding Liouville equation \p{eq:2.15} and the evolution equation \p{eq:2.14} for the
Jacobian (where $x_{i}=({\bf x}_{i},{\bf p}_{i})$) assume the form:
\begin{equation}\lab{eq:2.18}
{\partial{\cal D}\over\partial t}-\{H,{\cal D}\}=\sum_{1\leqslant
i\leqslant N} {\partial\over\partial p_{in}} \biggl({\cal
D}{\partial R\over\partial p_{in}}\biggr),
\end{equation}
\begin{equation}\lab{eq:2.19}
{\partial I\over\partial t}-\{H,I\}=\sum_{1\leqslant i\leqslant N}
{\partial\over\partial p_{in}} \biggl(I{\partial R\over\partial
p_{in}}\biggr),
\end{equation}
where $\{A,B\}$ is a Poisson bracket,
$$
\{A,B\}=\sum_{1\leqslant i\leqslant N}\biggl({\partial
A\over\partial x_{in}}{\partial B\over\partial p_{in}}-{\partial
A\over\partial p_{in}}{\partial B\over\partial x_{in}}\biggr).
$$
The operators $\Lambda(x)$, $\tilde\Lambda(x)$, which are defined
by \p{eq:2.6'}, \p{eq:2.16'}, become for the  case of Eqs.
\p{eq:2.1}:
\begin{equation}\lab{eq:2.20}
\Lambda(x)=\sum_{1\leqslant i\leqslant N}{p_{in}\over m}
{\partial\over\partial x_{in}}+\sum_{1\leqslant i,j\leqslant N}
F_{ij,n} {\partial\over\partial p_{in}},
\end{equation}
\begin{equation}\lab{eq:2.21}
\tilde\Lambda(x)=\sum_{1\leqslant i\leqslant N}{p_{in}\over m}
{\partial\over\partial x_{in}}+\sum_{1\leqslant i,j\leqslant N}
{\partial\over\partial p_{in}} F_{ij,n}.
\end{equation}

The generalized Liouville equation \p{eq:2.18} represents a basis for studying kinetics of systems
with dissipative interaction. The next step is to derive the corresponding generalized BBGKY
hierarchy, which is a starting point for derivation of kinetic equations. The $s$-particle
distribution function is defined by
\begin{equation} \lab{eq:2.22}
f_{s}(x_{1},...,x_{s};t)=V^{s}\int dx_{s+1}...dx_{N}{\cal
D}(x_{1},...,x_{N},t),
\end{equation}
where $V$ is the volume of the system and $x_{i}=({\bf x}_{i},{\bf p}_{i})$. Using the Liouville
equation \p{eq:2.18}, one now straightforwardly obtains the desired hierarchy:
\begin{equation} \lab{eq:2.24}
{\partial f_{s}\over\partial t}=-\tilde\Lambda_{s}f_{s}-\sum_{1\leqslant i\leqslant
s}{\partial\over\partial p_{in}}{1\over v}\int dx_{s+1}f_{s+1}F_{i\,s+1,n},
\end{equation}
where $\tilde\Lambda_{s}$ is given by Eq.(\ref{eq:2.21}) with $N=s$ and $1/v \equiv N/V$ is the
particle number density. The generalized BBGKY hierarchy \p{eq:2.24} reduces to the standard BBGKY
hierarchy when the dissipation function vanishes.

Our next goal is to solve the hierarchy \p{eq:2.24} employing a perturbative approach. To this end
we shall take the Bogolyubov principle of spatial correlation weakening \cite{Bogol} as a basis of
our consideration. In terms of the many-particle distribution functions $f_{s}(x_{1},...,x_{s};t)$,
this principle states that
\begin{equation}\lab{eq:2.26}
f_{s}(x_{1},...,x_{s},t)\xrightarrow[r\to\infty]{}
f_{s'}(x_{1}',...,x_{s'}',t)f_{s''}(x_{1}'',...,x_{s''}'',t),
\end{equation}
where two groups of $s'$ and $s''$ ($s=s'+s''$) particles are formed from $x_{1},...,x_{s}$ and $r$
is the minimal distance between the particles from different groups. The relation \p{eq:2.26} has a
simple physical meaning: the phase variables of particles are statistically independent at large
distance between particles. The property (\ref{eq:2.26}) of $f_{s}(x_{1},...,x_{s},t)$ holds in the
thermodynamic limit and it specifies a set of functions, in terms of which one should seek a
solution of the BBGKY hierarchy.

\section{ Kinetic stage of evolution} \label{weak}

The present section is devoted to a description of a kinetic stage of evolution for the dissipative
system under consideration. Following Bogolyubov, we assume his functional hypothesis as a basis of
our investigation \cite{Akh_Pel}. According to this hypothesis, for sufficiently large times,
many-particle distribution functions depend on time and initial distribution functions only through
one-particle distribution function \cite{Bog,Bogol}:
\begin{equation}\lab{eq:3.1}
f_{s}(x_{1},...,x_{s},t)\xrightarrow[t\gg\tau_{0}]{}
f_{s}(x_{1},...,x_{s};f(t)),
\end{equation}
where
$$
f_1(x_1,t)\xrightarrow[t\gg\tau_{0}]{}f(x_1,t).
$$
Here $f_{s}(x_{1},...,x_{s};f)$ are functionals of the one-particle distribution function, $\tau_0$
is a microscopic time, usually estimated as a collision time. One can consider the above functional
hypothesis \p{eq:3.1} as a generalization of the Chapman-Enskog approach to the derivation of
hydrodynamic equations proceeding from the Boltzmann equation.

According to the functional hypothesis (\ref{eq:3.1}), the functionals $f_{s}(x_{1},...,x_{s};f)$
are universal because they do not depend on initial conditions for the many-particle distribution
functions $f_{s}(x_{1},...,x_{s};t=0)$. These functionals can be calculated in a perturbative
approach. To illustrate the subsequent steps in this direction, we first study a more simple
perturbative approach for the case of weak interactions. The next section deals with the low
density limit, which is adequate to the situation of granular systems, in particular granular
fluids.

Following Eqs. \p{eq:2.24}, the single-particle distribution function satisfies the kinetic
equation of the form
\begin{equation}\lab{eq:3.5}
{\partial f(x_{1},t)\over\partial t}+{p_{1n}\over m}{\partial
f(x_{1},t)\over\partial x_{1n}}=L(x_{1}; f(t)),
\end{equation}
where the functional $L(x_{1},f)$ represents the generalized
collision integral,
\begin{equation}\lab{eq:3.6}
L(x_{1};f)=-{\partial\over\partial p_{1n}}{1\over v}\int
dx_{2}f_{2}(x_{1},x_{2};f)F_{12,n}.
\end{equation}
The Bogolyubov functional hypothesis \p{eq:3.1}, in self-evident shorthand notation, gives:
\begin{equation}\lab{eq:3.4}
{\partial f_{s}(f(t))\over\partial t}=\int dx{\left.{\delta
f_{s}(f)\over\delta f(x)}\right |}_{f\to f(t)}{\partial
f(x,t)\over\partial t},
\end{equation}
where $\delta f_{s}/\delta f$ denotes  a functional derivative. This relation and Eqs.\p{eq:2.24}
yield the following equation for the functional $f_s(f)$:
\begin{equation}\lab{eq:3.7}
-\int dx{\delta f_{s}(f)\over\delta f(x)}{p_n \over m} {\partial
f(x)\over\partial x_n}+\sum_{1\leqslant i\leqslant s}{ p_{in}\over
m}{\partial f_{s}(f)\over\partial x_{in}}=K_{s}(f)
\end{equation}
where we have introduced an auxiliary functional $K_s(f)$,
$$
K_{s}(f)=-\sum_{1\leqslant i,j\leqslant s}{\partial\over\partial
p_{in}}(f_{s}(f)F_{ij,n})-
$$
$$
-\sum_{1\leqslant i\leqslant s}{\partial\over\partial p_{in}}{1\over v}\int
dx_{s+1}f_{s+1}(f)F_{i\,s+1,n}-
$$
$$
-\int dx{\delta f_{s}(f)\over\delta f(x)}L(x;f).
$$

Next, following Bogolyubov again \cite{Bog,Bogol}, in order to obtain an unambiguous solution to
Eq. \p{eq:3.7} we need to formulate a asymptotical condition ("boundary condition") for the
functionals $f_{s}(f)$. This condition should reflect the principle of spatial correlation
weakening \p{eq:2.26} and it should be written taking into account the evolution of the system in
physical direction of time \cite{Bog,Bogol}. To this end, we introduce an auxiliary parameter
$\tau$, which has dimensions of time but does not represent physical time, and we use it in the
following manner:
$$
e^{-\tau \Lambda^0_s}f_s(x_1,...,x_s;f)=
$$
$$
=f_{s}\left({\bf x}_{1}-{{\bf p}_{1}\over m}\tau, {\bf
p}_{1},...,{\bf x}_{s}-{{\bf p}_{s}\over m}\tau, {\bf
p}_{s};f\right)\xrightarrow[\tau \sim +\infty]{}
$$
\begin{equation} \lab{eq:3.8}
\prod_{1\leqslant i\leqslant s}f\left({{\bf x}_{i}-{{\bf p}_{i}\over m}\tau}, {\bf
p}_{i}\right)=e^{-\tau \Lambda^0_s}\prod_{1\leqslant i\leqslant s}f(x_i).
\end{equation}
Here $e^{-t \Lambda^0_s}$ is the evolution operator of $s$ free particles and $\Lambda^0_s$ is
given by the first term in \p{eq:2.20} (see also \p{eq:2.21}):
\begin{equation} \lab{eq:3.9}
\Lambda_{s}^{0}=\sum_{1\leqslant i\leqslant s}{p_{in}\over m}{\partial\over\partial x_{in}}.
\end{equation}
The asymptotical condition \p{eq:3.8} can be written in a more compact form,
\begin{equation} \lab{eq:3.10}
\lim_{\tau \to +\infty}e^{-\tau
\Lambda^0_s}f_s(x_1,...,x_s;e^{\tau
\Lambda^0_1}f)=f^0_s(x_1,...,x_s;f),
\end{equation}
where
\begin{equation} \lab{eq:3.11}
f^0_s(x_1,...,x_s;f)\equiv \prod_{1\leqslant i\leqslant s}f(x_i).
\end{equation}

In order to solve Eqs.\p{eq:3.7} we recast them in the form
\begin{equation}\label{evol}
{\partial\over\partial\tau}e^{-\tau\Lambda_{s}^{0}}f_{s}(e^{\tau
\Lambda^0_1}f)=-e^{-\tau\Lambda_{s}^{0}}K_{s}(e^{\tau \Lambda^0_1}f).
\end{equation}
(The straightforward differentiation of (\ref{evol}) gives Eqs. (\ref{eq:3.7})). Upon integrating
Eq.~ \p{evol} over $\tau$ from $0$ to $+\infty$, and using the above boundary condition
\p{eq:3.10}, one obtains the following chain of integral equations for the distribution functions:
\begin{equation} \lab{eq:3.12}
f_{s}(f)=f^0_{s}(f)+\int_{0}^{+\infty}d\tau
e^{-\tau\Lambda_{s}^{0}} K_{s}(e^{\tau \Lambda^0_1}f).
\end{equation}
Equations~\p{eq:3.12} are solvable in a perturbative theory in weak interaction. In the leading
approximation in small parameter $\lambda$ ($F_{ij,n}\sim\lambda$), one obtains:
$$
f_{s}^{(0)}(f)=f_{s}^0(f),
$$
$$
L^{(1)}(x_{1};f)=-{1\over v}{\partial\over\partial
p_{1n}}f(x_{1})\int dx_{2}f(x_{2})F_{12,n}.
$$
This yields the following kinetic equation (see \p{eq:3.5}), correct to linear order in the
interaction strengh:
$$
{\partial f(x_{1})\over\partial t}+{p_{1n}\over m}{\partial f(x_{1})\over\partial x_{1n}}=
$$
\begin{equation}\lab{eq:3.13}
={1\over v}{\partial\over\partial p_{1n}}f(x_{1})\left(\int
dx_{2}f(x_{2}){\partial V_{12}\over\partial x_{1n}}+\int
dx_{2}f(x_{2}){\partial R_{12}\over\partial  p_{1n}}\right)
\end{equation}
In the absence of dissipative forces (i.e., in the case $R=0$) this kinetic equation reduces to a
kinetic equation of Vlasov type with a self-consistent field $U({\bf x}_{1})$ given by:
$$
U({\bf x}_{1})=\int d{\bf x}_{2}V_{12}\int d{\bf p}_{2}f({\bf
x}_{2},{\bf p}_{2}).
$$
Another simple case is that of spatial homogeneity (but in the presence of dissipation). Then, Eq.
\p{eq:3.13} transforms to:
\begin{equation}\lab{eq:3.14}
{\partial f({\bf p}_{1})\over\partial t}={1\over
v}{\partial\over\partial p_{1n}}f({\bf
p}_{1}){\partial\over\partial p_{1n}}\int d{\bf p}_{2}f({\bf
p}_{2})R_0({\bf p}_{12}),
\end{equation}
where
\begin{equation}\lab{eq:3.15}
R_0({\bf p})\equiv \int d{\bf x}R({\bf x};{\bf p}).
\end{equation}
When the dissipation function is given by \p{eq:2.4''}, one obtains from \p{eq:3.14}, \p{eq:3.15}
the following equations for the densities of energy, momentum, and particle number:
$$
{\partial\over\partial t}\int d{\bf p}{p^{2}\over 2m} f({\bf
p})=-{\gamma_0\over 2vm}\int d{\bf p}_{1} d{\bf p}_{2}f({\bf
p}_{1})f({\bf p}_{2}){\bf p}^2_{12}<0,
$$
$$
{\partial\over\partial t}\int d{\bf p}{\bf p}f({\bf p})=0, \quad
{\partial\over\partial t}\int d{\bf p}f({\bf p})=0,
$$
where
$$
\gamma_0\equiv\int d{\bf x}\gamma({\bf x}).
$$
We see that the system becomes cool (the kinetic energy decreases) during its time evolution as it
should be. The chain of integral equations \p{evol} allows of studying the higher order
approximations in interaction without any principal difficulties.

\section{ Kinetic equation for dilute gases with
dissipative interaction}\label{dilute}

The present section is devoted to the case of small density with arbitrary in strength short-range
interaction. In addition, we do not allow for the possibility of formation of complexes of
particles, so as to avoid the necessity to introduce additional distribution functions. However,
the Bogolyubov method can be applied here as well.

In principle, we can start from the chain of integral equations \p{eq:3.12}, as in the previous
section. However, this is not convenient as the density expansion would require the use of
nontrivial resummation techniques applied to the pertinent virial expansion (see, for example,
\cite{Akh_Pel}). Therefore, we choose to employ here an alternate approach, similar to those,
developed by Bogolyubov to derive the Boltzmann equation.

Clearly, the expansion of the distribution functions $f_{s}(x_{1},...x_{s};f)$ in Taylor functional
series in the one-particle distribution,  $f(x)$, is equivalent to a density expansion (in powers
of  $1/v$). Moreover, it is easy to see, on the basis of the structure of Eqs.~\p{eq:3.7}, that the
leading contribution to $f_{s}(f)$ is proportional to $f^s$. In accordance with this, it is
convenient to rewrite Eqs. \p{eq:3.7} in the form:
\begin{equation}\lab{eq:4.1}
-\int dx{\delta f_{s}(f)\over\delta f(x)}{p_n\over m}{\partial
f(x)\over\partial x_n}+\tilde\Lambda_{s}f_{s}(f)=Q_{s}(f),
\end{equation}
where we have introduced a new auxiliary functional $Q_{s}(f)$,
$$
Q_{s}(f)\equiv -\sum_{1\leqslant i\leqslant s}{\partial\over\partial p_{in}}{1\over v}\int
dx_{s+1}f_{s+1}(f)F_{i\,s+1,n}-
$$
$$
-\int dx{\delta f_{s}(f)\over\delta f(x)}L(x;f).
$$
The differential operator $\tilde{\Lambda}_{s}$ is defined by \p{eq:2.21} with $N=s$. The expansion
of the left-hand side of Eqs.~\p{eq:4.1} includes terms that are proportional to $f^{s}$ and higher
order contributions, whereas the right-hand side is, at least, of order $f^{s+1}$.

The next step is to formulate a boundary condition for (\ref{eq:4.1}) taking into account the
evolution of the system in physical direction of time. This boundary condition reflects the
principle of spatial correlation weakening. From (\ref{eq:2.17}), (\ref{eq:2.7}), (\ref{eq:2.6}) we
have
$$
e^{-\tau\tilde\Lambda_{s}}f_{s}(f)=I_s(x,\tau)e^{-\tau \Lambda_{s}} f_{s}(f)=
$$
$$
=I_s(x,\tau)f_{s}\left(X_1(-\tau,x),...,X_s(-\tau,x);f\right),
$$
where $I_{s}(x,\tau)$ denotes the Jacobian (\ref{eq:2.11}) for $s$-particle dynamics. The
application of the principle of spatial correlation weakening (\ref{eq:2.26}) now yields
$$
e^{-\tau\tilde\Lambda_{s}}f_{s}(f)\xrightarrow[\tau \sim +\infty]{}
$$
\begin{equation} \lab{eq:4.2}
I_s(x,\tau)\prod_{1\leqslant i\leqslant s}f\left(\vec X_i^*(x)-{\tau \over m}\vec P_i^*(x),\vec
P_i^*(x)\right).
\end{equation}
Here, following \cite{Bog,Bogol,Akh_Pel}, we have introduced the asymptotic coordinates and momenta
$X_{i}^{*}(x)=({\bf X}_{i}^{*}(x),$ ${\bf P}_{i}^{*}(x))$ ($i=1,...,s$),
$$
{\bf X}_i(t,x)\xrightarrow[t \sim -\infty]{}\vec X_i^*(x)+{t \over m}\vec P_i^*(x),
$$
\begin{equation}\lab{eq:4.3}
{\bf P}_i(t,x)\xrightarrow[t \sim -\infty]{}\vec P_i^*(x)
\end{equation}
($X_{i}(t,x)\equiv({\bf X}_{i}(t,x),{\bf P}_{i}(t,x))$). The asymptotic coordinates and momenta
$X_{i}^{*}(x)$ do exist, because, for long times in the past, the particles of the system with
interaction under consideration are in a state of free motion.
Now, according to (\ref{eq:4.2}), we need to find the limiting value of Jacobian, $I_{s}(x,t)$ as
$t \to +\infty$. Making use the definition (\ref{eq:2.11}), one obtains
$$
I_s(x,t)=\bigg\vert{\partial X(-t,x)\over\partial x}\bigg\vert=\bigg\vert{\partial
X(-t,x)\over\partial X^{*}(x)}\bigg\vert\bigg\vert{\partial X^{*}(x)\over\partial x}\bigg\vert,
$$
whence, exploiting \p{eq:4.3}, we have
\begin{equation}\lab{eq:4.9}
I_s(x,t)\xrightarrow[t \to+\infty]{}I_s^{*}(x)\equiv\bigg\vert{\partial X^{*}(x)\over\partial
x}\bigg\vert.
\end{equation}
As a result, \p{eq:4.2} can be written in the final form
\begin{equation}\lab{eq:4.10}
 \lim_{\tau \to
+\infty}e^{-\tau\tilde\Lambda_{s}}f_{s}(x;e^{\tau\Lambda_1^0}f)=f_s^{(s)}(x;f),
\end{equation}
where
\begin{equation}\lab{eq:4.11}
f_s^{(s)}(x;f)\equiv I_s^*(x)\prod_{1\leqslant i\leqslant s}f( X_i^*(x))
\end{equation}
(above, in \p{eq:3.10}, \p{eq:3.11}, we have used a more detailed notation $(x_{1},...,x_{s})\equiv
x$). To solve Eqs. \p{eq:4.1} taking into account the obtained boundary condition \p{eq:4.10}, we
rewrite it as follows:
\begin{equation}\label{eq:4.12'}
{\partial\over\partial\tau}e^{-\tau\tilde\Lambda_{s}}f_{s}(e^{\tau
\Lambda^0_1}f)=-e^{-\tau\tilde\Lambda_{s}}Q_{s}(e^{\tau \Lambda^0_1}f).
\end{equation}
(The straightforward differentiation of \p{eq:4.12'} gives Eqs. \p{eq:4.1}). Integration of this
chain of equations over $\tau$ from $0$ to $+\infty$ yields the following integral equations for
the many-particle distribution functions:
\begin{equation} \lab{eq:4.12''}
f_{s}(f)=f_{s}^{(s)}(f)+\int_{0}^{+\infty}d\tau
e^{-\tau\tilde\Lambda_{s}}Q_{s}(e^{\tau \Lambda^0_1}f),
\end{equation}
Equations \p{eq:4.12''} are solvable in a perturbative approach in density. Similar integral
equations were obtained in \cite{Akh_Pel} for Hamiltonian systems. The difference between both
equations consists in the presence of the asymptotical value of Jacobian $I_s^*(x)$ in Eqs.
\p{eq:4.11}.

In the leading order in density, Eqs. \p{eq:4.12''} give the two-particle distribution function,
which is proportional to the squared density,
\begin{equation} \lab{eq:4.12}
f_{2}^{(2)}(x_{1},x_{2};f)=I_2^{*}(x_{1},x_{2})\prod_{1\leqslant
i\leqslant 2}f({\bf X}_{i}^{*}(x_{1},x_{2}),{\bf
P}_{i}^{*}(x_{1},x_{2})).
\end{equation}
Next, using  \p{eq:3.5} and \p{eq:3.6}, one obtains the  following kinetic equation:
\begin{equation}\lab{eq:4.13}
{\partial f(x_{1},t)\over\partial t}+{p_{1n}\over m}{\partial
f(x_{1},t)\over\partial x_{1n}}=L^{(2)}(x_{1};f(t)),
\end{equation}
where the collision integral $L^{(2)}(x_{1};f)$ is determined by
\begin{equation} \lab{eq:4.14}
L^{(2)}(x_{1};f)=-{1\over v}{\partial\over\partial p_{1n}}\int dx_{2}f_{2}^{(2)}(x_{1},x_{2};f)
F_{12,n}.
\end{equation}
The kinetic equation \p{eq:4.13} is a generalization of the
Boltzmann kinetic equation to non-Hamiltonian systems. The
collision integral \p{eq:4.14} is written in the Bogolyubov form
and expressed through the asymptotic coordinates and momenta,  and
the  Jacobian corresponding to two-particle dynamics.

In the case of weakly nonuniform states, when the gradients of the one one-particle distribution
function $f(\vec x,\vec p)$ are small, the collision integral \p{eq:4.14} can be further
simplified. For these states, the range $r_{0}$ of the interpartical forces is small compared to
the characteristic scale of inhomogeneity $a$, $r_{0}\ll a$, i.e. in comparison to those distances
over which the one-particle distribution function $f(\vec x,\vec p)$ changes substantially. Also,
we take into account that
\begin{equation}\lab{eq:4.15}
|{\bf X}_{i}^{*}(x_1,x_2)-{\bf x}_{i}|\sim r_{0}, \quad {\bf P}_{i}^{*}(x_{1},x_{2})\equiv {\bf
P}_{i}^{*}({\bf x}_{21},{\bf p}_{1},{\bf p}_{2}),
\end{equation}
($i=1,2$). Following Eq.~(\ref{eq:4.12}) these asymptotic
coordinates and momenta of the two particles problem determine the
two-particle distribution function $f_{2}^{(2,0)}(x_{1},x_{2};f)$
to  second order in the particle density and  zeroth  order in the
gradients. Using \p{eq:4.1} one obtains:
$$
{{p_{12,n}}\over m}{\partial
f_{2}^{(2,0)}(x_{1},x_{2};f)\over\partial{
x_{1n}}}+{\partial\over\partial
p_{1n}}(f_{2}^{(2,0)}(x_{1},x_{2};f)F_{12,n})+
$$
$$
+{\partial\over\partial p_{2n}}(f_{2}^{(2,0)}(x_{1},x_{2};f)
F_{21,n})=0.
$$
Integration of this equation over $x_{2}$ leads, using \p{eq:4.14}, to an expression for the
collision integral $L^{(2,0)}(x;f)$,
\begin{equation} \lab{eq:4.15'}
L^{(2,0)}(x_{1};f)={1\over v}\int dx_{2}{{p_{21,n}}\over m}{\partial
f_{2}^{(2,0)}(x_{1},x_{2};f)\over\partial x_{2n}},
\end{equation}
where
$$
f_{2}^{(2,0)}(x_{1},x_{2};f)=
$$
\begin{equation} \lab{eq:4.16}
=I_2^{*}(x_{1},x_{2})f({\bf x}_{1},{\bf
P}_{1}^{*}(x_{1},x_{2}))f({\bf x}_{1},{\bf
P}_{2}^{*}(x_{1},x_{2}))
\end{equation}
(see \p{eq:4.12}, \p{eq:4.15}).

Now, we evaluate integral over ${\bf x}_{2}$ ($dx_{2}=d{\bf x}_{2}d{\bf p}_{2}$) in \p{eq:4.15'}.
The integration can be replaced by an integration over the difference ${\bf x}_{21}$ (see
\p{eq:4.12}, \p{eq:4.15}). In performing the integral over ${\bf x}_{21}$, we employ cylindrical
coordinates $z$, $b$ and $\varphi$ with the origin at the point ${\bf x}_{1}$ and the $z$-axis
directed along the vector ${\bf p}_{21}$:
$$
L^{(2,0)}(x_{1};f)={1\over v}\int d{\bf
p}_{2}\int_{0}^{2\pi}d\varphi\int_{0}^{\infty}db\,\,b\,\,{|{\bf p}_{21}|\over m}\times
$$
\begin{equation} \lab{eq:4.17}
\times f_{2}^{(2,0)}(x_{1},x_{2}; f)\vert_{z=-\infty}^{z=+\infty}
\end{equation}
where $f_{2}^{(2,0)}(x_{1},x_{2};f)$ is given by \p{eq:4.16}. The asymptotic momenta
$$
{\bf P}_{i}^{*}({\bf x}_{\perp},z,{\bf p}_{1},{\bf p}_{2})\equiv {\bf P}_{i}^{*}({\bf x}_{21},{\bf
p}_{1},{\bf p}_{2}),
$$
which determine $f_{2}^{(2,0)}(x_{1},x_{2};f)$, have the following properties:
$$
{\bf P}_{i}^{*}({\bf x}_{\perp},z,{\bf p}_{1},{\bf
p}_{2},)\vert_{z\to +\infty}={\bf p}^{\prime}_{i}(b,\varphi,{\bf
p}_{1},{\bf p}_{2}),
$$
\begin{equation} \lab{eq:4.19}
{\bf P}_{i}^{*}({\bf x}_{\perp},z,{\bf p}_{1},{\bf
p}_{2})\vert_{z\to-\infty}={\bf p}_{i},
\end{equation}
where ${\bf x}_{\perp}=({\bf x}_{21})_{\perp}=(b,\varphi)$, $z={\bf x}_{21} {\bf p}_{21}/|{\bf
p}_{21}|$. Indeed, according to \p{eq:4.3}, ${\bf P}_i^*(x_1,x_2)$ are the momenta of two particles
at the moment of time $t=-\infty$, if at $t=0$ they have phase variables $\vec x_1,\vec p_1, \vec
x_2, \vec p_2$. Then, the relationship
$$
\left|(\vec x_1+{\vec p_1 \over m}t)-(\vec x_2+{\vec p_2 \over m}t)\right|=|\vec x_{12}|+{z \over
t_0}t+O(t^2)
$$
($t_0 \equiv {|\vec x_{21}|m / |\vec p_{21}|}$), which is valid for small $t$, shows the following:
when $z>0$, the collision of particles precedes the moment $t=0$, whereas when $z<0$, the collision
occurs after $t=0$. This reasoning explains the relations \p{eq:4.19}, where ${\bf
p}^{\prime}_{1}(b,\varphi,{\bf p}_{1},{\bf p}_{2})$, ${\bf p}^{\prime}_{2}(b,\varphi,{\bf
p}_{1},{\bf p}_{2})$ are the momenta of particles before the collision (precollisional  momenta)
after which the particles have momenta ${\bf p}_1$, ${\bf p}_2$.

With these observation we can now find the following expression
for the generalized Boltzmann collision integral determined by
\p{eq:4.16} and \p{eq:4.17}:
$$
L^{(2,0)}(x_{1};f)={1\over v}\int d{\bf p}_{2}\int_{0}^{2\pi}d\varphi\int_{0}^{\infty}db\,b\,{|{\bf
p}_{21}|\over m}\times
$$
\begin{equation}\lab{eq:4.20}
\times\{I_2'({\bf x}_{\perp},{\bf p}_{1},{\bf p}_{2})f({\bf
x}_{1},{\bf p}_{1}')f({\bf x}_{1},{\bf p}_{2}')- f({\bf
x}_{1},{\bf p}_{1})f({\bf x}_{1},{\bf p}_{2})\},
\end{equation}
where
\begin{equation}\lab{eq:4.21}
I_2'({\bf x}_{\perp},{\bf p}_{1},{\bf
p}_{2})=I^{*}(x_{1},x_{2})\vert_{z=+\infty}.
\end{equation}
The calculation of the collision integral $L^{(2,0)}(x_{1};f)$
(much like the calculation of the two-particle distribution
function)  involves only the solution of the two-particle
dynamics. This is elaborated in the subsection that appears
immediately below.

\section{Dissipative dynamics}\label{disdin}

\subsection{Relative motion in two-particle dynamics}

For the Hamiltonian systems the two-particle problem is reduced to study of relative motion of the
particles. The same situation takes place in the presence of dissipative forces.

To obtain this result and some its consequences, let us consider the equations of motion for two
particles in the presence of dissipative forces. According to Eqs.~\p{eq:2.1}-\p{eq:2.4}, these
equations have the form
\begin{equation}\lab{eq:5.1}
m\dot{\bf x}_{1}={\bf p}_{1}, \qquad m\dot{\bf x}_{2}={\bf p}_{2},
\end{equation}
\begin{equation}\lab{eq:5.1'}
\dot{\bf p}_{1}={\bf F}({\bf x}_{12};\,{\bf p}_{12}), \qquad
\dot{\bf p}_{2}=-{\bf F}({\bf x}_{12};\,{\bf p}_{12}),
\end{equation}
where
$$
F_n({\bf x};{\bf p})=-{{\partial V({\bf x})}\over\partial
x_{n}}-{{\partial R({\bf x};{\bf p})}\over\partial p_{n}}.
$$
Let us introduce the following new phase variables ${\bf x}$, ${\bf p}$, ${\bf x_c}$, ${\bf p_c}$
with clear meaning:
$$
{\bf x}={\bf x}_{1}-{\bf x}_{2}, \qquad {\bf p}={{{\bf p}_{1}-{\bf
p}_{2}}\over 2},
$$
\begin{equation}\lab{eq:5.2}
{\bf x_c}={{{\bf x}_{1}+{\bf x}_{2}}\over 2}, \qquad {\bf
p_c}={\bf p}_{1}+{\bf p}_{2}.
\end{equation}
Then, the equations of motion \p{eq:5.1}, \p{eq:5.1'} are separated into equations for the center
of mass and relative motion:
\begin{equation}\lab{eq:5.5}
2m\dot{\bf x}_c={\bf p_c}, \qquad \dot{\bf p}_c=0,
\end{equation}
\begin{equation}\lab{eq:5.5'}
m\dot{\bf x}=2{\bf p}, \qquad \dot{\bf p}={\bf F}({\bf x},2{\bf
p}).
\end{equation}

Let ${\bf x}(t,{\bf x},{\bf p})$, ${\bf p}(t,{\bf x},{\bf p})$ be a solution of Eqs.~\p{eq:5.5'}
with initial conditions ${\bf x}$, ${\bf p}$. According to \p{eq:4.3}, \p{eq:5.2}, the
corresponding asymptotic coordinates and momenta are given by
$$
{\bf p}(t,{\bf x},{\bf p})\xrightarrow[t \sim -\infty]{}{\bf
p}^{*}({\bf x},{\bf p}),
$$
\begin{equation}\lab{eq:5.6}
{\bf x}(t,{\bf x},{\bf p})\xrightarrow[t\sim -\infty]{}{\bf
x}^{*}({\bf x},{\bf p})+{2t\over m}{\bf p}^{*}({\bf x},{\bf p}),
\end{equation}
where
$$
\vec p^*(\vec x,\vec p)\equiv {1\over 2}\left(\vec P_1^*(x_1,x_2)-\vec P_2^*(x_1,x_2)\right)
$$
\begin{equation} \label{eq:5.6'}
\vec x^*(\vec x,\vec p)\equiv \vec X_1^*(x_1,x_2)-\vec X_2^*(x_1,x_2).
\end{equation}
Integrating Eqs. \p{eq:5.5}, we can also come to the following identities:
$$
\vec p_1+\vec p_2=\vec P^*_1(x_1,x_2)+\vec P^*_2(x_1,x_2),
$$
\begin{equation}\label{eq:5.6''}
\vec x_1+\vec x_2=\vec X^*_1(x_1,x_2)+\vec X^*_2(x_1,x_2).
\end{equation}

The comparison of Eqs. \p{eq:5.6}-\p{eq:5.6''} enables to express the asymptotic coordinates and
momenta ${\bf {X}}_{i}^{*}(x_{1}, x_{2})$ and ${\bf P}_{i}^{*}(x_{1},x_{2})$ through the functions
$\vec x^*(\vec x,\vec p)$, $\vec p^*(\vec x,\vec p)$
$$
{\bf P}_{1}^{*}(x_{1},x_{2})={{\bf p_c}\over 2}+{\bf p}^{*}({\bf
x},{\bf p}), \,\,{\bf P}_{2}^{*}(x_{1},x_{2})={{\bf p_c}\over
2}-{\bf p}^{*}({\bf x},{\bf p}),
$$
\begin{equation}\lab{eq:5.10}
{\bf X}_{1}^{*}(x_{1},x_{2})={\bf x_c}+{1\over 2}{\bf x}^{*}({\bf
x},{\bf p}), \,\, {\bf X}_{2}^{*}(x_{1},x_{2})={\bf x_c}-{1\over
2}{\bf x}^{*}({\bf x},{\bf p}).
\end{equation}
We can see that the calculation of asymptotic phase variables for the two-particle dynamics is
reduced to the calculation of asymptotic phase variables for the relative motion.

Next, consider the  Jacobian $I_2(x_{1},x_{2},t)$ that corresponds to the dynamics of two particles
and determines the collision integral \p{eq:4.20}. According to Eq.\p{eq:2.19}, this Jacobian
satisfies the following equation:
$$
{\partial I_2\over\partial t}-\{H_{2},I_2\}={\partial\over\partial p_{1n}}\biggl(I_2{{\partial
R}\over\partial p_{1n}}\biggr)+{\partial\over\partial p_{2n}}\biggl(I_2{\partial R\over\partial
p_{2n}}\biggr),
$$
where $H_{2}$ is the  two-particle Hamiltonian (see \p{eq:2.2} for $N=2$). Changing the independent
variables in this equation to ${\bf x}$, ${\bf p}$, ${\bf x}_c$, ${\bf p}_c$ (see \p{eq:5.2}), one
finds:
$$
{\partial I_2\over\partial t}+{p_{cn}\over 2m}{\partial
I_2\over\partial x_{cn}}+{2p_n\over m}{\partial I_2\over\partial
x_n}-{\partial V\over\partial x_n}{\partial I_2\over\partial
p_n}={1\over 2}{\partial\over\partial p_n}\left(I_2{\partial
R\over\partial p_n}\right).
$$
Since $I_2(x_1,x_2,0)=1$, it follows from the latter equation that the Jacobian does not depend on
${\bf x}_c$ and ${\bf p}_c$, i.e. $I_2(x_1,x_2,t)\equiv I_2(\vec x, \vec p,t)$. The Jacobian
$I_2(\vec x, \vec p,t)$ satisfies equation
\begin{equation}\lab{eq:5.11}
{\partial I_2\over\partial t}-\{h,I_2\}={1\over
2}{\partial\over\partial p_n}\biggl(I_2{\partial R\over\partial
p_n}\biggr), \qquad I_2\vert_{t=0}=1,
\end{equation}
where
\begin{equation}\lab{eq:5.12}
\{h,I_2\}={\partial h\over\partial x_n}{\partial I_2\over\partial p_n}-{\partial h\over\partial
p_{n}}{\partial I_2\over\partial x_n}, \qquad h\equiv{p^{2}\over m}+V({\bf x}).
\end{equation}
Next,  introduce the Jacobian $\tilde I_2({\bf x},{\bf p},t)$ corresponding to the dynamics defined
by  \p{eq:5.5'}. It can be easily seen, using Eqs. \p{eq:2.19}, \p{eq:5.5'}, that this Jacobian
satisfies the same equation and initial condition as $I_2({\bf x},{\bf p},t)$ (see Eqs.
\p{eq:5.11}, \p{eq:5.12}). Therefore, these two Jacobians are equal to each other,
\begin{equation}\lab{eq:5.12'}
I_2({\bf x},{\bf p},t)=\tilde{I}_2({\bf x},{\bf p},t).
\end{equation}
Taking into account this result and the definition \p{eq:4.9} of the limiting Jacobian, we obtain
\begin{equation}\lab{eq:5.12''}
I^*_2(x_1,x_2)=I^*_2(\vec x,\vec p)={\partial (\vec x^*,\vec
p^*)\over\partial (\vec x,\vec p)}.
\end{equation}

\subsection{Two-particle dynamics with weak
dissipation}

The present  section is devoted to the study of the case of weak dissipation, namely the kinetics
for which it is sufficient to consider only the linear order in an expansion of collision integral
in powers of the dissipation function. This case is similar to the corresponding expansion in
powers of the degree of inelasticity \cite{Sela_Gold}.

In accordance with Eqs. \p{eq:4.9}, \p{eq:4.19}-\p{eq:4.21}, in order to derive a kinetic equation
in the case of weak dissipation, we have to calculate the asymptotic coordinates ${\vec
X}^*_i(x_1,x_2)$ and momenta ${\vec P}^*_i(x_1,x_2)$ $(i=1,2)$ in a perturbative approach in the
dissipation function $R(\vec x,\vec p)$. However, in the previous sub-section, we have showed that
it is sufficient to find the asymptotic coordinates ${\vec x}^*(\vec x, \vec p)$ and momenta ${\vec
p}^*(\vec x, \vec p)$ for relative motion. This motion is described by the solution ${\vec
x}(t,x)\equiv{\vec x}(t,\vec x, \vec p)$, ${\vec p}(t,x)\equiv{\vec p}(t,\vec x, \vec p)$  of Eqs.
\p{eq:5.5'}, which can be written in the form
\begin{equation}\lab{eq:6.1}
{\vec x}(t,\vec x, \vec p)=e^{t(\lambda_0+\lambda_1)}\vec x,\quad {\vec p}(t,\vec x, \vec
p)=e^{t(\lambda_0+\lambda_1)}\vec p,
\end{equation}
where the operators $\lambda_0$, $\lambda_1$ have the following structure:
\begin{equation}\lab{eq:6.2}
\lambda_0\equiv {2p_n \over m}{\partial \over
\partial x_n}-{\partial V(\vec x)\over \partial x_n}{\partial
\over \partial p_n}, \quad \lambda_1\equiv -{1\over 2}{\partial
R(\vec x, 2\vec p)\over \partial p_n}{\partial \over \partial p_n}
\end{equation}
(see Eqs. \p{eq:2.5}-\p{eq:2.6'}).

In the sequel, while calculating ${\vec x}(t,\vec x, \vec p)$, ${\vec p}(t,\vec x, \vec p)$ we
shall consider $\lambda_1$ as a small perturbation. The unperturbed relative motion is expressed as
\begin{equation}\lab{eq:6.3}
{\bf p}^{(0)}(t,\vec x, \vec p)=e^{t\lambda_0}{\bf p}, \quad {\bf
x}^{(0)}(t,\vec x, \vec p)=e^{t\lambda_0}{\bf x}.
\end{equation}
Since the operators $\lambda_0$ and $\lambda_1$ do not commute, we
use the following well known expansion:
\begin{equation}\lab{eq:6.4}
e^{t(\lambda_0+\lambda_1)}=e^{t\lambda_0}+\int_{0}^{t}dt'
e^{t'\lambda_0}\lambda_1e^{(t-t') \lambda_0}+...
\end{equation}
This formula, in conjunction with  \p{eq:6.1}-\p{eq:6.3}, gives
$$
{\bf p}(t,x)={\bf p}^{(0)}(t,x)+\int_{0}^{t}dt'\left.\lambda_1\,{\bf p}^{(0)}(t-t',x)\right|_{x\to
x^{(0)}(t',x)},
$$
$$
{\bf x}(t,x)={\bf x}^{(0)}(t,x)+\int_{0}^{t}dt'\left.\lambda_1\,{\bf x}^{(0)}(t-t',x)\right|_{x\to
x^{(0)}(t',x)}
$$
$(x^{(0)}(t,x)\equiv ({\vec x}^{(0)}(t,x), {\vec p}^{(0)}(t,x))$. Using \p{eq:5.6}, it is easy to
find the asymptotic momentum ${\bf p}^{*}(x)$ and coordinate ${\bf x}^{*}(x)$ by letting $t$ to
$-\infty$:
$$
{\bf p}^{*}(x)={\bf p}^{*(0)}(x)-\int_{-\infty}^{0}dt\left.\lambda_1\,{\bf p}^{*(0)}(x)\right
|_{x\to x^{(0)}(t,x)},
$$
$$
{\bf x}^{*}(x)={\bf x}^{*(0)}(x)-\int_{-\infty}^{0}dt\,\lambda_1\,\bigg\{ {\bf x}^{*(0)}(x)-
$$
\begin{equation} \lab{eq:6.9}
-{2t\over m}{\bf p}^{*(0)}(x)\bigg\}\bigg\vert_{x\to x^{(0)}(t,x)}.
\end{equation}

Consider now the Jacobian $I_2(x,t)$ in the linear order in the dissipation function $R$. To this
order in $R$, it follows from Eq. \p{eq:5.11} (with $I_{2}(x,t)=1$ at $R=0$) that
$$
{\partial I_2\over\partial t}+\lambda_0 I_2={1\over 2}{\partial^{2}R({\bf x},2{\bf p})\over\partial
p_n\partial p_n},
$$
whence
$$
I_{2}(x,t)=1+{1\over 2}\int_{-t}^{0}dt'\left.{\partial^{2}R({\bf x},2{\bf p})\over\partial
p_n\partial p_n}\right |_{x\to x^{(0)}(t',x)}.
$$
According to Eq. \p{eq:4.9}, the asymptotic value of the Jacobian $I_2(x,t)$ is given by the
formula:
\begin{equation}\lab{eq:6.10}
I^{*}_2(x)=1+{1\over 2}\int_{-\infty}^{0}dt\left.{\partial^{2}R({\bf x},2{\bf p})\over\partial
p_n\partial p_n}\right |_{x\to x^{(0)}(t,x)}.
\end{equation}

\section{Kinetic equation in the weak
dissipation approximation}\label{weakdisskin}

In this section we study the kinetic equation \p{eq:3.5} with the collision integral \p{eq:4.20} in
the weak dissipation approximation. In the spatially homogeneous case and in the linear
approximation in $R$, the collision integral Eq. \p{eq:4.20} assumes the form:
\begin{equation}\lab{eq:6.11}
L^{(2,0)}({\bf p}_{1};f)=L_{0}^{(2,0)}({\bf p}_{1};f)+L_{1}^{(2,0)}({\bf p}_{1};f),
\end{equation}
where $L_{0}^{(2,0)}({\bf p}_{1},f)$ is the Boltzmann collision
integral, which accounts only for the reversible (potential)
interactions:
$$
L_{0}^{(2,0)}({\bf p}_{1};f)={1\over v}\int d{\bf
p}_{2}\int_{0}^{2\pi}d\varphi\int_{0}^{\infty}db\,b{{|{\bf p}_{21}|}\over m}\times
$$
\begin{equation}\lab{eq:6.11'}
\times\{f({\bf p}_{10}')f({\bf p}_{20}')-f({\bf p}_{1})f({\bf
p}_{2})\}
\end{equation}
(${\vec p}_{i0}'\equiv {\vec p_i'}^{(0)}$ are the Boltzmann precollisional momenta). The second
term in \p{eq:6.11} is a correction to the Boltzmann collision integral, which accounts for
dissipation to linear order in $R$:
$$
L_{1}^{(2,0)}({\bf p}_{1};f)={1\over v}\int d{\bf
p}_{2}\int_{0}^{2\pi}d\varphi\int_{0}^{\infty}db\,b{{|{\bf p}_{21}|}\over m}\times
$$
$$
\times\delta[I'({\bf x}_{\perp},{\bf p}_{1},{\bf p}_{2})f({\bf
p}_{1}')f({\bf p}_{2}')],
$$
where ${\bf x}_{\perp}=({\bf x}_{21})_{\perp}=(b,\varphi)$ (see section 4). It is clear that $I'=1$
for the reversible dynamics (when $R=0$). Thus, taking into account that $\delta{\bf
p}_{1}'=-\delta{\bf p}_{2}'$ (see \p{eq:4.19}, \p{eq:5.10}) and using \p{eq:6.9}, \p{eq:6.10} for
the asymptotic  values of the momentum and Jacobian, we find
$$
L_{1}^{(2,0)}({\bf p}_{1};f)={1\over v}\int d{\bf
p}_{2}\int_{0}^{2\pi}d\varphi\int_{0}^{\infty}db\,b{{|{\bf p}_{21}|}\over m}\times
$$
$$
\times\bigg\{a({\bf x}_{\perp},{\bf p}) +b_n({\bf x}_{\perp},{\bf p})\biggl({\partial\over\partial
p_{1n}}-{\partial\over\partial p_{2n}}\biggr)\bigg\}\times
$$
\begin{equation}\lab{eq:6.12}
\times\left. f({\bf p}_{1})f({\bf p}_{2})\right|_{{\bf p}_{1},{\bf
p}_{2}\to{{\bf p}_{10}^{\prime},{{\bf p}_{20}^{\prime}}}},
\end{equation}
where
\begin{equation}\lab{eq:6.13}
a({\bf x}_{\perp},{\bf p})={1\over
2}\int_{-\infty}^{0}dt\left.{\partial^{2} R\over{\partial p_n
\partial p_n}} \right |_{x\to x^{(0)}(t,x),\,z\to+\infty},
\end{equation}
\begin{equation}\lab{eq:6.14}
b_n({\bf x}_{\perp},{\bf p})={1\over 2}\int_{-\infty}^{0}dt\left.{\partial R\over\partial
p_{m}}{\partial p_{n}^{*(0)}\over\partial p_{m}} \right |_{x\to x^{(0)}(t,x),\,z\to+\infty}
\end{equation}
($z=({\bf x}{\bf p})/|{\bf p}|$, see section 4).

Next we wish to evaluate  \p{eq:6.12} for the  case in which the
particle interactions vanish ($V=0$).  In this case the  solution
of the equations of motion assumes  of the form
$$
{\bf x}^{(0)}(t,x)={\bf x}+{2t\over m}{\bf p}
$$
and, as expected, the asymptotic  momentum and space coordinate
coincide with their respective initial values, ${\bf
p}^{*(0)}={\bf p}$, ${\bf x}^{*(0)}={\bf x}$. Noting that
$$
\int_{-\infty}^{0}dtg\biggl({\bf x}+{2t\over m}{\bf
p}\biggr)=\int_{-\infty}^{0}dtg\biggl({\bf x}_{\perp},z+{2t\over
m}p\biggr)=
$$
$$
={m\over 2p}\int_{-\infty}^{z}dz'g({\bf x}_{\perp},z')
$$
is valid for an arbitrary function $g({\bf x})=g({\bf x}_{\perp},z)$ (the $z$-axis of the
cylindrical coordinates we employ  is chosen to point in the direction of  ${\bf p}$), one finds,
using \p{eq:6.13}, \p{eq:6.14}, that:
$$
a({\bf x}_{\perp},{\bf p})={m\over
4p}\int_{-\infty}^{\infty}dz{\partial^{2} R(\vec x
_{\perp},z,2\vec p)\over\partial p_n\partial p_n},
$$
$$
b_n({\bf x}_{\perp},{\bf p})={m\over
4p}\int_{-\infty}^{\infty}dz{\partial R(\vec x _{\perp},z,2\vec
p)\over\partial p_n}.
$$
Finally, upon substituting these expressions into \p{eq:6.12} and noting  that ${\bf p}_{10}'={\bf
p}_{1}$, ${\bf p}_{20}'={\bf p}_{2}$ for $V=0$, one obtains
\begin{equation}\lab{eq:6.15}
L_{1}^{(2,0)}({\bf p}_{1};f)={1\over v}{\partial\over\partial
p_{1n}}f({\bf p}_{1}){\partial\over\partial p_{1n}}\int d{\bf
p}_{2}f({\bf p}_{2})R_0({\bf p}_{12}),
\end{equation}
where $R_0({\bf p})$ is defined by Eq. \p{eq:3.15}. Formula
\p{eq:6.15} coincides with \p{eq:3.14} obtained within the weak
interaction approximation.

In conclusion of this section we briefly concern the question of the evolution of the system
described by the kinetic equation
\begin{equation}\lab{eq:6.15'}
{\partial f({\bf p},t)\over\partial t}=L_{0}^{(2,0)}( {\bf
p};f(t))+L_{1}^{(2,0)}({\bf p};f(t)).
\end{equation}
Let $\tau_{r}$ be the relaxation time defined by the usual Boltzmann term in Eq. \p{eq:6.15'}. In
the absence of dissipative interaction this relaxation leads to the Maxwellian distribution for
$f({\bf p},t)$. However, in the presence of small dissipative interaction described by the second
term in Eq. \p{eq:6.15'}, we shall observe a weak relaxation of temperature of the system, i.e. the
homogeneous cooling state (see, for example, \cite{IGreview}). The description of this state can be
based on the functional hypothesis of the form
\begin{equation}\lab{eq:6.16}
f(\vec p,t)\xrightarrow[t\gg\tau_{r}]{}f(\vec p,\varepsilon(t)),
\end{equation}
where the asymptotic value of the energy density $\varepsilon(t)$ is defined by
$$
\int d\vec p f(\vec p,t){\vec p^2 \over 2m}\xrightarrow
[t\gg\tau_{r}]{}\varepsilon(t).
$$
This functional hypothesis results to the equation for $\varepsilon(t)$,
$$
{\partial \varepsilon(t)\over \partial t}=L(\varepsilon(t))
$$
with the following right-hand side:
$$
L(\varepsilon)\equiv \int d\vec p {\vec p^2 \over 2m}L_{1}^{(2,0)}({\bf p}_{1};f(\varepsilon))
$$
(the Boltzmann collision integral $L_{0}^{(2,0)}({\bf p}_{1};f)$ does not contribute to
$L(\varepsilon)$).

The distribution function $f(\varepsilon)$, according to Eq. \p{eq:6.15'} and the functional
hypothesis \p{eq:6.16}, satisfies the equation
\begin{equation}\lab{eq:6.17}
{\partial f(\varepsilon)\over \partial\varepsilon }L(\varepsilon)=
L_{0}^{(2,0)}( {\bf p};f(\varepsilon))+L_{1}^{(2,0)}({\bf
p};f(\varepsilon)).
\end{equation}
This equation is solvable in a perturbative approach in powers of the dissipation function. We
shall not discuss here the study of the homogeneous cooling state based on the obtained equations.
This can be done similar to those theory developed for spatially nonuniform states in
\cite{Sela_Gold,Gol'd_Sok}. Finally, we note that in another terminology, the sketched theory is
the application of the Chapman-Enskog method to the solution of the kinetic equation \p{eq:6.15'}.

\section{Connection to the Boltzmann equation for inelastic
rigid spheres }\label{spheres}

In this section we compare our kinetic equation \p{eq:4.20} with that obtained by considering a
system of rigid particles experiencing instantaneous inelastic collisions characterized by a fixed
coefficient of normal restitution (see, e.g. \cite{Sela_Gold,Gol'd_Sok}):
$$
L'(\vec p_1;f)={d^{2}\over mv}\int_{ \vec k \vec p_{12}>0}d{\bf p}_{2}\int d^{2}{\bf k}(\vec k {\bf
p}_{12})\times
$$
\begin{equation}\lab{eq:7.1}
\times \{{1\over\varepsilon^{2}}f({\bf p}_{1}^{\prime})f({\bf p}_{2}^{\prime})-f({\bf p}_{1})f({\bf
p}_{2})\},
\end{equation}
where ${\bf k}$ is a unit vector pointing  from the center of sphere $1$ to that of sphere $2$ at
the moment of contact ($d^{2}{\bf k}=\sin{\theta}d\theta d\varphi$; the polar axis $z$ is directed
along ${\bf p}_{21}={\bf p}_{2}-{\bf p}_{1}$), $d$ is the diameter of a sphere, and $\varepsilon$
is the coefficient of normal restitution. Using the identity
$$
\int_{0}^{2\pi}d\varphi\int_{0}^{\infty}b\,db|{\bf p}_{12}|...=d^{2}\int_{{\bf k}{\bf
p}_{21}>0}d^{2}{\bf k}({\bf k}{\bf p}_{21})...
$$
($b=d\sin\theta$, $0\leqslant\theta\leqslant\pi/2$), one obtains
from \p{eq:7.1}:
$$
L'(\vec p_1;f)={1\over v}\int d{\bf p}_{2}\int_{0}^{2\pi}d\varphi\int_{0}^{\infty}db\ b\ {{|{\bf
p}_{21}|}\over m}\times
$$
\begin{equation}\lab{eq:7.2}
\times\bigg\{{1\over\varepsilon^{2}}f({\bf p}_{1}')f({\bf p}_{2}')-f({\bf p}_{1})f({\bf
p}_{2})\bigg\}.
\end{equation}
Here ${\bf p}_{1}'$, ${\bf p}_{2}'$ are the precollisional momenta are determined by
\begin{equation}\lab{eq:7.3}
{\bf p}_{1}'={\bf p}_{1}+{{1+\varepsilon}\over\varepsilon}{\bf
k}({\bf p}{\bf k}), \quad {\bf p}_{2}'={\bf
p}_{2}-{{1+\varepsilon}\over\varepsilon}{\bf k}({\bf p}{\bf k}),
\end{equation}
where ${\bf p}=({\bf p}_1-{\bf p}_2)/2$. The collision integral \p{eq:4.20} is determined by the
asymptotic ($t\to -\infty$) values of the momenta, coordinates, and Jacobian, which specify the
two-particle dynamics. In terms of the relative momentum ${\bf p}$, the collision law \p{eq:7.3}
can be written as
\begin{equation}\lab{eq:7.4}
{\bf p}'={\bf p}+{{1+\varepsilon}\over\varepsilon}{\bf k}({\bf
pk}).
\end{equation}
Within the framework of the formalism developed in this article, one needs to know the relation
between the asymptotic values  of coordinate ${\bf x}'$ and the initial coordinate ${\bf x}$. We
establish this relation in the terms orf the relative coordinate $\vec x=\vec x_1- \vec x_2$ as
follows:
\begin{equation}\lab{eq:7.4a}
{\bf x}^{\prime}={\bf x}+{{1+\varepsilon}\over\varepsilon}{\bf k}({\bf x k}).
\end{equation}
Using the above  ${\bf x}^{\prime}$ (we assume ${\bf k}={\rm
const}$) we obtain the Jacobian:
$$
{\partial({\bf x}',{\bf p}')\over\partial({\bf x},{\bf
p})}={1\over\varepsilon^{2}}.
$$
Substitution of this Jacobian into the collision integral \p{eq:4.20} gives the collision integral
\p{eq:7.2}. Therefore, when  \p{eq:7.3}, \p{eq:7.4} are satisfied, the collision integrals
\p{eq:4.20}, \p{eq:7.1} coincide, as they should.

\section{Conclusion}
\label{conclusion}

We have shown that the Bogolyubov method of derivation of kinetic
equations can be applied to dissipative many-body systems with the
corresponding modifications. In the case of inelastically
colliding hard spheres we reproduce the inelastic Boltzmann
equation. The reader may be justified  in asking whether yet
another formulation is needed to study dissipative systems. We
believe that the answer is that given the difficulties encountered
by other approaches, in particular the problems  emanating from
the lack of scale separation in granular systems, it is
advantageous to consider a powerful approach such as that of
Bogolyubov. The application of this approach to dense systems, for
instance, would not only serve to complement the results obtained
by using the Enskog corrected Boltzmann equation, but may also
enable the study of systems (such as binary granular mixtures)
where a naive application of the Enskog-Boltzmann equation has
been shown to be invalid even in the framework of elastically
interacting particles \cite{mixtures}. Much like any other
approach to many-body systems, the present one is not directly
useful: perturbative expansions need to be implemented to obtain
physically significant results. However, as the formulation is
rather different from e.g.,  those directly based on the Boltzmann
equation or its ring corrections, one may be able to study
hitherto inaccessible cases (or limits), e.g. when gradients are
large (typical of granular systems) or many-body contacts are of
importance. Whether the present approach will indeed provide
useful results for these and other cases of dissipative systems
remains to be seen. \vspace{0,3cm}

\noindent

{\bf Acknowledgments}

\vspace{0,2cm}

The authors gratefully acknowledge the useful discussions with Yu.V. Slyusarenko





\end{document}